\documentclass[preprint]{aastex63}
\usepackage{amsmath,amssymb,CJK,longtable}
\usepackage{booktabs}  

\received{-}
\revised{-}
\accepted{-}
\submitjournal{PSJ}

\shorttitle{Emission of (3200) Phaethon At 1~AU}
\shortauthors{Ye et al.}

\begin{document}
\begin{CJK*}{UTF8}{gbsn}

\title{A Deep Search for Emission From ``Rock Comet'' (3200) Phaethon At 1~AU}

\correspondingauthor{Quanzhi Ye}
\email{qye@umd.edu}

\author[0000-0002-4838-7676]{Quanzhi Ye (叶泉志)}
\affiliation{Department of Astronomy, University of Maryland, College Park, MD 20742, USA}

\author[0000-0003-2781-6897]{Matthew M. Knight}
\affiliation{Department of Physics, United States Naval Academy, 572C Holloway Rd, Annapolis, MD 21402, USA}
\affiliation{Department of Astronomy, University of Maryland, College Park, MD 20742, USA}

\author[0000-0002-6702-7676]{Michael S. P. Kelley}
\affiliation{Department of Astronomy, University of Maryland, College Park, MD 20742, USA}

\author[0000-0001-6765-6336]{Nicholas A. Moskovitz}
\affiliation{Lowell Observatory, 1400 W Mars Hill Road, Flagstaff, AZ 86001, USA}

\author[0000-0002-7600-4652]{Annika Gustafsson}
\affiliation{Department of Astronomy and Planetary Science, P.O. Box 6010, Northern Arizona University, Flagstaff, AZ 86011, USA}

\author{David Schleicher}
\affiliation{Lowell Observatory, 1400 W Mars Hill Road, Flagstaff, AZ 86001, USA}



\begin{abstract}

We present a deep imaging and spectroscopic search for emission from (3200) Phaethon, a large near-Earth asteroid that appears to be the parent of the strong Geminid meteoroid stream, using the 4.3~m Lowell Discovery Telescope. Observations were conducted on 2017 December 14--18 when Phaethon passed only 0.07~au from the Earth. We determine the $3\sigma$ upper level of dust and CN production rates to be 0.007--0.2~$\mathrm{kg~s^{-1}}$ and $2.3\times10^{22}~\mathrm{molecule~s^{-1}}$ through narrowband imaging. A search in broadband images taken through the SDSS {\it r'} filter shows no 100-m-class fragments in Phaethon's vicinity. A deeper, but star-contaminated search also shows no sign of fragments down to 15~m. Optical spectroscopy of Phaethon and comet C/2017 O1 (ASASSN) as comparison confirms the absence of cometary emission lines from Phaethon and yields $3\sigma$ upper levels of CN, C$_2$ and C$_3$ of $\sim10^{24}$--$10^{25}~\mathrm{molecule~s^{-1}}$, 2 orders of magnitude higher than the CN constraint placed by narrowband imaging, due to the much narrower on-sky aperture of the spectrographic slit. We show that narrowband imaging could provide an efficient way to look for weak gas emission from near-extinct bodies near the Earth, though these observations require careful interpretation. Assuming Phaethon's behavior is unchanged, our analysis shows that the {\it DESTINY$^+$} mission, currently planning to explore Phaethon in 2026, may not be able to directly detect a gas coma.

\end{abstract}



\section{Introduction} \label{sec:intro}

(3200) Phaethon, discovered with the Infrared Astronomical Satellite in 1983 \citep{Gibson1983, Green1983}, is one of the very few near-Earth asteroids that have been directly or indirectly observed to exhibit mass-loss activities \citep{Jewitt2015, Ye2018b}. It has been linked with the Geminid meteoroid stream which produces the strongest annual meteor shower on the Earth in modern times \citep{Whipple1983, Williams1993}. Most meteor showers have been associated with unambiguous comets that usually become active at 3--5~au from the Sun due to the sublimation of water ice \citep{Meech2004}. With an aphelion distance of 2.4~au, Phaethon's orbit is entirely within the water sublimation regime; however, numerous observations of Phaethon in the past four decades have so far failed to find signs of water-driven activity. Various deep imaging and spectroscopic searches have established the upper bound of emission levels of $\sim10^{23}~\mathrm{molecule~s^{-1}}$ for CN \citep{McFadden1985, Chamberlin1996}, and $\sim10^{-2}~\mathrm{kg~s^{-1}}$ for dust \citep{Hsieh2005, Tabeshian2019}, scaled to $\sim1$~au from the Sun. Phaethon has so far only exhibited weak activity near its perihelion at $\sim0.14$~au, presumably due to thermal fracturing caused by the extreme environmental temperature at this distance to the Sun \citep{Jewitt2010, Li2013, Hui2017}. This activity is several orders of magnitude too low to supply the Geminid meteoroid stream.

While Phaethon has not been observed to show mass-loss activity except for near perihelion, a few near-Earth asteroids have been found to be active at much more moderate distance to the Sun (e.g. $\sim1$~au). (3552) Don Quixote, for example, has displayed recurrent activity from 1.23--2.72 au \citep{Mommert2014,Mommert2020}; (196256) 2003~EH$_1$, associated with the Quadrantid meteor shower, might have been active when its perihelion distance was $\sim1$~au \citep{Jenniskens2004, Abedin2015, Gulashina2017}. If Phaethon exhibits activity at $\gg0.14$~au, it will point to a mechanism different than the one that drives its near-perihelion activity. This could also have significant implication for JAXA's DESTINY$^+$ mission\footnote{An acronym of ``Demonstration and Experiment of Space Technology for INterplanetary voYage, Phaethon fLyby and dUst Science Phaethon fLyby with reUSable probe''.}, which will flyby Phaethon in $\sim2026$ at a heliocentric distance of  $\sim0.9$~au \citep{Sarli2018, Arai2019}, and therefore would potentially encounter dust ejecta released at such distance.

Phaethon passed only 0.07~au from the Earth on 2017 December 16, making its closest approach to the Earth between 1974 and 2093. This provided an exceptional opportunity to study Phaethon up close, and has facilitated studies from optical to infrared and radio wavelength \citep[e.g.][and others]{Devogele2018, Kareta2018, Kim2018, Ito2018, Schmidt2018, Kartashova2019, Taylor2019}. Here we present a deep imaging and spectroscopic search for Phaethon's emission conducted during this close encounter.

\section{Observation and Data Reduction} \label{sec:obs}

We observed Phaethon with the 4.3~m Lowell Discovery Telescope (LDT; formerly known as the Discovery Channel Telescope, DCT) from 2017 December 14 to 18, near the time of its closest approach to the Earth. At the time of the observation, Phaethon was approaching its next perihelion on 2018 February 11. Imaging and spectroscopic observations were made using the Large Monolithic Imager \citep[LMI;][]{Massey2013} and the DeVeny spectrograph \citep{Bida2014}, respectively, with the observational circumstances summarized in Table~\ref{tbl:obs}. All observations were made with the telescope tracking at Phaethon's rate of motion. Data were obtained with both LMI and Deveny on 2017 December 16, but were not as constraining as other nights due to the weather, so are not reported here. Observations with the Near-Infrared High Throughput Spectrograph (NIHTS) were also made and will be reported in a separate paper (Gustafsson et al. submitted).

LDT has an instrument cube which supports the simultaneous mounting of five instruments. Fold mirrors allow rapid (${\lesssim}2$~min) switches between instruments, making it possible to quickly transition back and forth between instruments depending on science goals and changing atmospheric conditions. A dichroic transmits optical light to LMI while sending infrared light to NIHTS, allowing simultaneous observations of visible imaging and near-infrared spectroscopy. The dichroic has $\gtrsim$90\% transmission from roughly 4000--7000~{\AA} and results in reduced sensitivity around a central, unvignetted field of view, and renders some of the LMI filters (e.g. {\it CN}) unusable due to the decreased wavelength range. It was used for a subset of images as discussed below.

\begin{deluxetable}{cccccccccccc}
\tablecaption{Summary of imaging and spectroscopic observations of Phaethon presented in this work. The abbreviations of the symbols are $r_\mathrm{H}$ -- heliocentric distance, $\varDelta$ -- geocentric distance, and $\alpha$ -- phase angle. Sections marked with \dag~indicate the use of dichroic (see main text). \label{tbl:obs}}
\tabletypesize{\scriptsize}
\setlength{\tabcolsep}{0.05in}
\tablehead{
\colhead{Date (UT)} & \colhead{Purpose} & \colhead{Section} & \colhead{Instrument} & \colhead{Filter} &  \colhead{$r_\mathrm{H}$} & \colhead{$\varDelta$} & \colhead{$\alpha$} & \colhead{Airmass} & \colhead{Exposure} & \colhead{Cond.\tablenotemark{\scriptsize{a}}} &
\colhead{FWHM\tablenotemark{\scriptsize{b}}} \\
\colhead{} &
\colhead{} &
\colhead{} &
\colhead{} &
\colhead{} &
\colhead{(au)} &
\colhead{(au)} &
} 
\startdata
2017 Dec 14 & Dust & {{ \protect\S~\ref{sect:obs:bc} }}, {{ \protect\S~\ref{sect:dust} }} & LMI & {\it BC} & 1.06 & 0.09 & $32^\circ$ & 1.01--1.16 & (5~s)$\times518$ (512 used)$^\dag$ & IC & $1.0''$--$1.4''$ \\
2017 Dec 14 & Dust & {{ \protect\S~\ref{sect:obs:bc} }}, {{ \protect\S~\ref{sect:dust} }} & LMI & {\it BC} & 1.06 & 0.09 & $32^\circ$ & 1.01--1.16 & (5~s)$\times518$ (512 used)$^\dag$ & IC & $1.0''$--$1.4''$ \\
'' & CN gas & {{ \protect\S~\ref{sect:obs:cn} }}, {{ \protect\S~\ref{sect:cn} }} & LMI & {\it CN} & 1.06 & 0.09 & $32^\circ$ & 1.05--1.06 & (30~s)$\times12$ & P & $1.0''$ \\
2017 Dec 15 & Gas & {{ \protect\S~\ref{sect:obs:spec} }}, {{ \protect\S~\ref{sect:spec} }} & DeVeny & - & 1.04 & 0.08 & $41^\circ$ & 1.01--1.03 & (60~s)$\times29$ & P & $1.9''$--$2.0''$ \\
2017 Dec 17 & Fragment & {{ \protect\S~\ref{sect:obs:r} }}, {{ \protect\S~\ref{sect:frag} }} & LMI & {\it r'} & 1.01 & 0.07 & $68^\circ$ & 1.01--1.16 & (3~s)$\times25$, (300~s)$\times24$ (8 used) & CL & $1.3''$--$1.7''$ \\
2017 Dec 18 & Dust, fragment & {{ \protect\S~\ref{sect:obs:r} }}, {{ \protect\S~\ref{sect:frag} }} & LMI & {\it r'} & 0.99 & 0.07 & $81^\circ$--$83^\circ$ & 1.06--2.65 & (0.5~s)$\times238^\dag$, (0.7~s)$\times86^\dag$, (1~s)$\times90$\tablenotemark{\scriptsize{c}}$^{, \dag}$ & IC & $1.2''$--$1.8''$ \\
\enddata
\tablenotetext{a}{Weather conditions: CL -- intermittent clouds; IC -- intermittent cirrus; P -- photometric}
\tablenotetext{b}{Full-width-half-maximum (FWHM) of field stars.}
\tablenotetext{c}{Only some are used -- see \S~\ref{sect:obs:r}.}
\end{deluxetable}

\subsection{Imaging}

LMI has an unvignetted field-of-view (FOV) of $12.3'\times12.3'$ without dichroic or approximately $6.4'\times4.1'$ with dichroic and a pixel scale of $0.36''$ after an on-chip $3\times3$ binning. We observed Phaethon in three filters, {\it BC}, {\it CN}, and {\it r'}, to be described in the following paragraphs. We also observed with NIHTS on December 14 and 18, so some images from both nights used the dichroic. The {\it r'} images from December 18 were acquired with NIHTS aligned at the parallactic angle, therefore image orientation varied throughout that night. Additionally, NIHTS took very short exposures and was dithered 5$''$ in the parallactic direction between each exposure to facilitate thermal background subtraction. Given the differing readout and exposure times for each instrument, synchronization of image cadence would have significantly reduced the on-sky efficiency of NIHTS. We thus elected to observe with NIHTS as the primary instrument, resulting in occasional LMI images that needed to be discarded due to telescope motion mid-exposure.

\begin{figure*}[!htb]
\centering
\includegraphics[width=\textwidth]{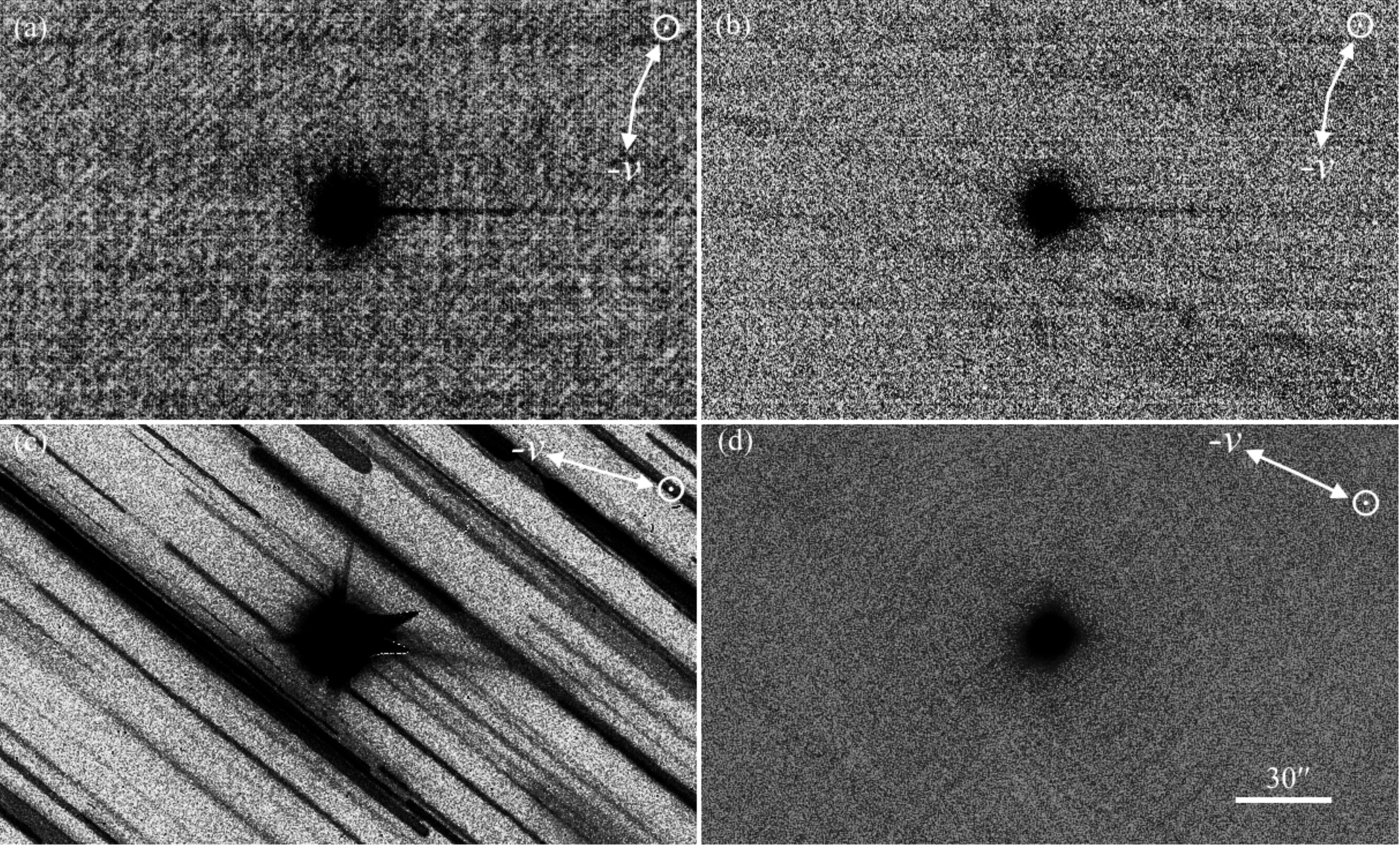}
\caption{Median-combined images in different filters: (a) $512\times5$ s in {\it BC} taken 2017 December 14; (b) $12\times30$ s in {\it CN}, 2017 December 14; (c) $8\times300$ s in {\it r'}, 2017 December 17; and (d) a total integrated exposure of 120.7 s in {\it r'} from 206 images taken 2017 December 18. North is up and east is to the left. The tail-like feature towards 3 o'clock in (a)--(c) is a detector readout artifact. All images are stretched using the histogram equalization model (which redistributes pixel values such that the output image has a flat distribution of intensities) and are color-inverted. Also shown on the upper-right corner of each panel are the comet-sun vector (arrow to $\odot$) and negative velocity vector (arrow to $-v$).}
\label{fig:img}
\end{figure*}

Several calibration stars were also observed to facilitate data calibration. All images were bias-subtracted and flat-field corrected with flats in the appropriate configuration (with or without the dichroic) using {\tt ccdproc} \citep{Craig2015}. Images in each filter set were then aligned and median combined into one final ``stacked'' image per night for analysis using {\tt AstroImageJ} \citep{Collins2017} and {\tt PHOTOMETRYPIPELINE} \citep{Mommert2017}, shown as Figure~\ref{fig:img}.

\subsubsection{Blue Continuum $BC$}
\label{sect:obs:bc}

We obtained 518 images through the ``blue'' narrowband filter for dust continuum, {\it BC}, each in 5~sec exposure. A total of 6 images were discarded due to trailing that resulted from the motion of the telescope during NIHTS operations. The {\it BC} filter, defined in the HB\footnote{Stands for ``Hale--Bopp'' in reference to C/1995 O1 (Hale--Bopp), but the official name of the filter system is HB \citep[][\S~1]{Farnham2000}.} narrowband filter system~\citep{Farnham2000}, has a central wavelength 4450~\AA~and a bandpass width of 67~\AA, and is nearly devoid of cometary gaseous species, most notably C$_2$ and C$_3$ around this wavelength. As such, the {\it BC} images are well-suited for the search of dust emission from Phaethon.

We observed two flux standards at different airmasses (HD 37112 at 1.31, HD 72526 at 2.67), as well as a point-spread-function (PSF) standard star HD 278835 (airmass 1.05, similar to Phaethon's airmass). The flux standards were used to determine the photometric calibration constants, while the PSF comparison star was used for comparison with Phaethon's brightness profile (\S~\ref{sect:dust}).

\subsubsection{Narrowband CN}
\label{sect:obs:cn}

We obtained 30 frames though the narrowband {\it CN} filter with 30~sec exposure for each frame. These observations were made without the dichroic. The {\it CN} filter, defined in the same HB system, has a central wavelength of 3870~\AA~and a bandpass width of 62~\AA. The filter can efficiently isolate CN gas, as the band is dominated by CN emission. The few ions presented within this bandpass are weak, minor species such as emissions from certain transition levels from C$_3$ and CH. However, given the anticipated extremely low (if any) activity from Phaethon, we conclude that the fluxes from these ions should be much weaker than the CN emission and are therefore negligible. Similar to the {\it BC} observation, we also observed the same flux standards (HD 37112 at an airmass of 1.30, HD 72526 at 2.74) in order to determine the photometric constants. We did not observe any solar analog star in {\it CN}, since the purpose of the {\it CN} observation is to facilitate a simple search of gas coma which does not depend on removal of the solar continuum.

\subsubsection{Broadband Sloan {\it r'}}
\label{sect:obs:r}

Broadband Sloan {\it r'} images were obtained over two nights (2017 December 17 and 18). This standard filter has a central wavelength at 6200~\AA~and a bandpass width of 1200~\AA. We adopted a different exposure strategy on each night: for the first night, we conducted a number of long, 5~min exposures, interspersed with some short, 3~sec exposures, with Phaethon purposely saturated in the longer exposures to facilitate a deep search for smaller fragments in its vicinity; for the second night, we conducted a large number of very short exposures varying from 0.5 to 1~sec, with the same goal of searching for fragments in Phaethon's vicinity. The unsaturated images from the second night would also be used for the search of dust emission from Phaethon, complementing the search with the {\it BC} images (\S~\ref{sect:obs:bc}). Compared to the {\it BC} images, the {\it r'}-band images are susceptible to weak C$_2$ and NH$_2$ emissions present in its bandpass, though we note that detection of any emission would be scientifically interesting.

The untrailed images were photometrically calibrated using the field stars and the Pan-STARRS DR1 catalog \citep{Chambers2016}, using the \texttt{PHOTOMETRYPIPELINE} package \citep{Mommert2017}. The 5~min exposures taken on 2017 December 17, with background stars significantly trailed due to the high apparent motion of Phaethon, were paired with untrailed 3~sec images that were taken closest in time. The photometric constants of each 5~min image were then scaled from the paired 3~sec image considering the difference in the integration time.

Both nights were affected by clouds, though the situation on the second night (2017 December 18) was significantly better than the first (2017 December 17). For the images on the first night, we visually inspected all images and discarded 2/3 of the images that were significantly impacted by the clouds. The cloud-affected images from the second night passed the initial visual check, but were later found to introduce a weak ($\sim10\%$) gradient to the resultant stacked image (i.e. one side of the image is slightly brighter than the other). Some images were also taken at high airmasses (up to 2.65) that may have contributed to the issue. Therefore, we rejected the images with photometric zero points beyond the $3\sigma$ range from the median value, as well as those with airmass greater than 2, totaling 208 out of 414 images taken on that night. The following analysis is based on the stacked image generated using this filtered image set.

\subsection{Spectroscopy}
\label{sect:obs:spec}

Spectroscopic observations of Phaethon were made with LDT's DeVeny optical spectrograph, which was formerly known as the Kitt Peak White Spectrograph before being transferred to Lowell Observatory in 1998. DeVeny was first installed at the 1.8~m Perkins Telescope, and was moved and upgraded for operation on LDT in 2015 \citep{Bida2014}. The upgraded camera on DeVeny has an unbinned pixel scale of $0.34''$. We observed flux standard Feige 24 and solar analog HD 28099 at similar airmasses as the Phaethon observations (1.17 for Feige 24, 1.06 for HD 28099) to enable absolute flux calibration and removal of reflected solar continuum of Phaethon. Additionally, we also observed a bright unambiguous comet, C/2017 O1 (ASASSN), just before the Phaethon observation, to facilitate direct comparison between Phaethon and a comet. C/ASASSN was observed at an airmass of 1.63, hence we observed an additional solar analog SA 93-101 at an airmass of 1.69 for solar continuum removal of the C/ASASSN spectrum. We used a $1''\times150''$ slit and a grating of 150 l/mm throughout our observations, providing a resolving power of $\sim500$ (with a dispersion of 4.3~\AA/pixel at the blaze wavelength of 5000~\AA) and a spectral coverage of 3200~\AA~to 1~\micron. All DeVeny observations were made at the parallactic angle.

We performed standard image reduction using the Image Reduction and Analysis Facility, or {\tt IRAF} \citep{Tody1986, Tody1993}, including bias subtraction, flat-fielding correction, spectrum extraction, wavelength and absolute flux calibration. We used different aperture lengths to extract the spectrum of different objects: $3.4''$ (10 pixels, or $\pm5$ pixels from the nucleus; equivalent to 200~km at Phaethon, 2700~km at C/ASASSN) for flux standards and solar analogs, and $68''$ (200 pixels, or $\pm100$ pixels from the nucleus; equivalent to 4500~km at Phaethon, 53,000~km at C/ASASSN) for Phaethon and C/ASASSN. The aperture for Phaethon and C/ASASSN is much larger since the goal is to capture as much signal of the large coma as possible. To ensure that the large aperture does not dilute the possible signal from the coma, we also process the data in parallel using a smaller aperture length of $17''$ ($\pm25$~pixels from the nucleus, or 1000~km at Phaethon, 13,000~km at C/ASASSN), which showed no improvement over the ones processed using the $68''$ aperture. Spectra were extracted using the defined apertures centered on the target. Sky backgrounds, determined using a pair of $34''$-long apertures immediately outside the target aperture and symmetric to the target, were subtracted from each spectral image. The analyses, to be described in the following sections, were performed on these on-target, background-subtracted spectra.

Following an experimental technique discussed in \citet{Chamberlin1996}, we also perform an experiment and extracted two off-target spectra from the Phaethon images, with extraction apertures covering the range between the Phaethon signal and the edge of the image, symmetric with respect to the Phaethon signal. The purpose of the off-target spectra was to test an alternative way to determine the gas emission from Phaethon, as it was free from the contamination from Phaethon itself and could theoretically provide a wider on-sky aperture due to the extended nature of the hypothetical CN coma. However, similar to \citet{Chamberlin1996}'s finding, we discovered that the constraint derived from the off-target spectra was worse than the on-target spectrum. Hence, we only perform our analysis on the on-target spectrum.

\section{Results}

\subsection{Dust}
\label{sect:dust}

\subsubsection{Radial Brightness Profile}
\label{sect:dust:profile}

\begin{figure*}[!htb]
\centering
\includegraphics[width=\textwidth]{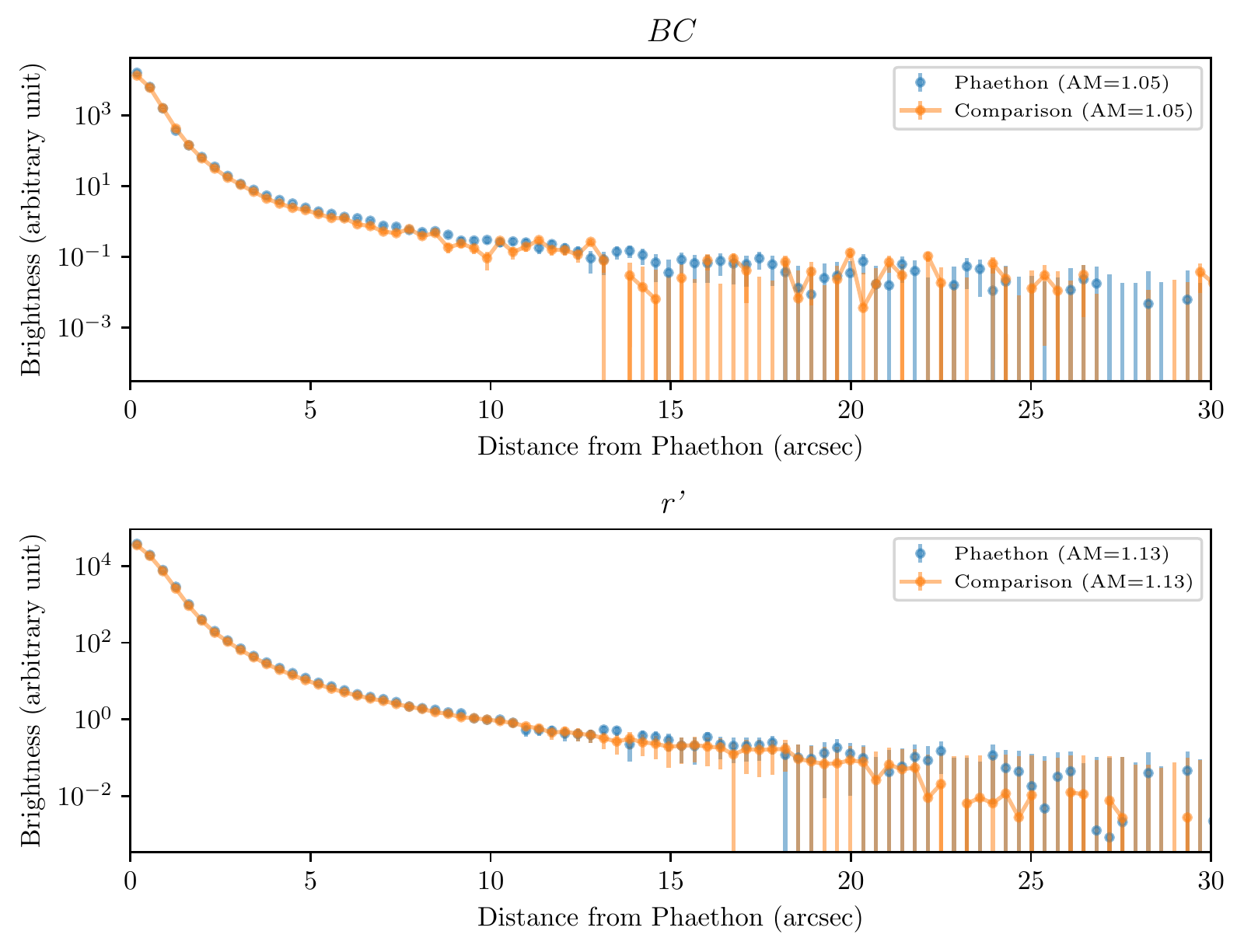}
\caption{Radial brightness profiles of Phaethon and comparison star taken using {\it BC} filter (top panel) and {\it r'} filter (bottom panel) under the same airmasses.}
\label{fig:profile}
\end{figure*}

We first looked for extendedness of Phaethon that indicated the presence of a dust coma. This was done by comparing Phaethon's brightness profile to a comparison star. For {\it BC} images, we used PSF comparison star HD 278835, observed at a similar airmass as the Phaethon images, as the comparison star. For {\it r'} images, since no PSF comparison star was observed, we chose a field star with similar brightness, TYC 1720-1205-1 ($V=10.65\pm0.05$), as the comparison star, though we note that Phaethon's apparent motion was so high that any field star was only visible for a small number of frames. Uncertainties are computed assuming that they are dominated by Poisson statistics. Despite the very high rate of motion ($\sim$2100{\arcsec}~hr$^{-1}$), our exposure times were short enough that stars trailed for much less than the seeing ($1''$--$2''$ for those nights). The result, shown as Figure~\ref{fig:profile}, shows no extendedness in Phaethon's brightness profile, in line with the negative results from previous analyses \citep[e.g.][]{Hsieh2005, Tabeshian2019}.

\subsubsection{Aperture Photometry}
\label{sect:dust:aper}

To derive a more quantitative constraint on Phaethon's dust production, we performed aperture photometry on Phaethon. The size of the hypothetical dust coma, $l$, can be approximated by the turnaround distance of the sunward extent of the coma \citep[e.g.][]{Baum1992, Farnham2005, Hui2017}:

\begin{equation}
\label{eq:coma-size}
    l = \frac{v_\mathrm{ej}^2 r_\mathrm{H}^2}{2 \beta G M_\odot \varDelta \sin{\alpha}} \propto \frac{v_\mathrm{ej}^2}{\beta \sin{\alpha}}
\end{equation}

\noindent where $v_\mathrm{ej}$ is the grain ejection speed, $\varDelta$ and $r_\mathrm{H}$ are the geocentric and heliocentric distance of Phaethon at the time of the observation, $\beta=5.74\times10^{-4} Q_\mathrm{pr}/(\rho_\mathrm{d} a)$ \citep[c.f.][]{Burns1979} is the ratio of radiation pressure and gravitational force acting on the grain (where $Q_\mathrm{pr}\approx1$ is the scattering efficiency, $\rho_\mathrm{d}$ and $a$ are the bulk density and radius of the grain), $G$ is the universal gravitational constant, $M_\odot$ is the solar mass, and $\alpha$ is the phase angle. Here we see that $l$ is dependent on $v_\mathrm{ej}$ which, in turn, is dependent on the underlying activity mechanism which is unknown for Phaethon. Hence, we consider two possibilities:

\begin{enumerate}
    \item If the activity is driven by volatile sublimation like typical comets, the ejection speed $v_\mathrm{ej}$ varies as a function of $\sqrt{\beta}$ \citep{Whipple1950}, therefore $l$ is constant for all $\beta$ given the relation in Equation~\ref{eq:coma-size}. The exact value of $v_\mathrm{ej}$ depends on the heliocentric distance and intrinsic properties of the comet: the $v_\mathrm{ej}$ of typical water-ice-driven comets can be approximated by the \citet{Whipple1950} model, while that of some weakly active comets can be an order of magnitude lower \citep{Ye2016}. Considering these two different scenarios, we derive a theoretical apparent coma radius of $0.2'$--$5'$ for Phaethon during its 2017 encounter (corresponding to a $v_\mathrm{ej}$ between $\sim20$--$100~\mathrm{m~s^{-1}}$).
    \item It has been suggested that the modern day activity of Phaethon exhibited around its perihelion may be caused by thermal fracturing instead of water-ice sublimation \citep[e.g.][]{Jewitt2010, Li2013, Hui2017}. The ejection speed driven by such mechanism is not high, possibly approaching the gravitational escape limit \citep{Ye2019}, making $v_\mathrm{ej}$ dependent on $\beta$. Assuming gravitational escape ejection, the apparent coma radius of Phaethon during its 2017 encounter is $<20''$ for grains that are millimeter-sized or smaller.
\end{enumerate}

\begin{figure}[!htb]
\centering
\includegraphics[width=0.7\textwidth]{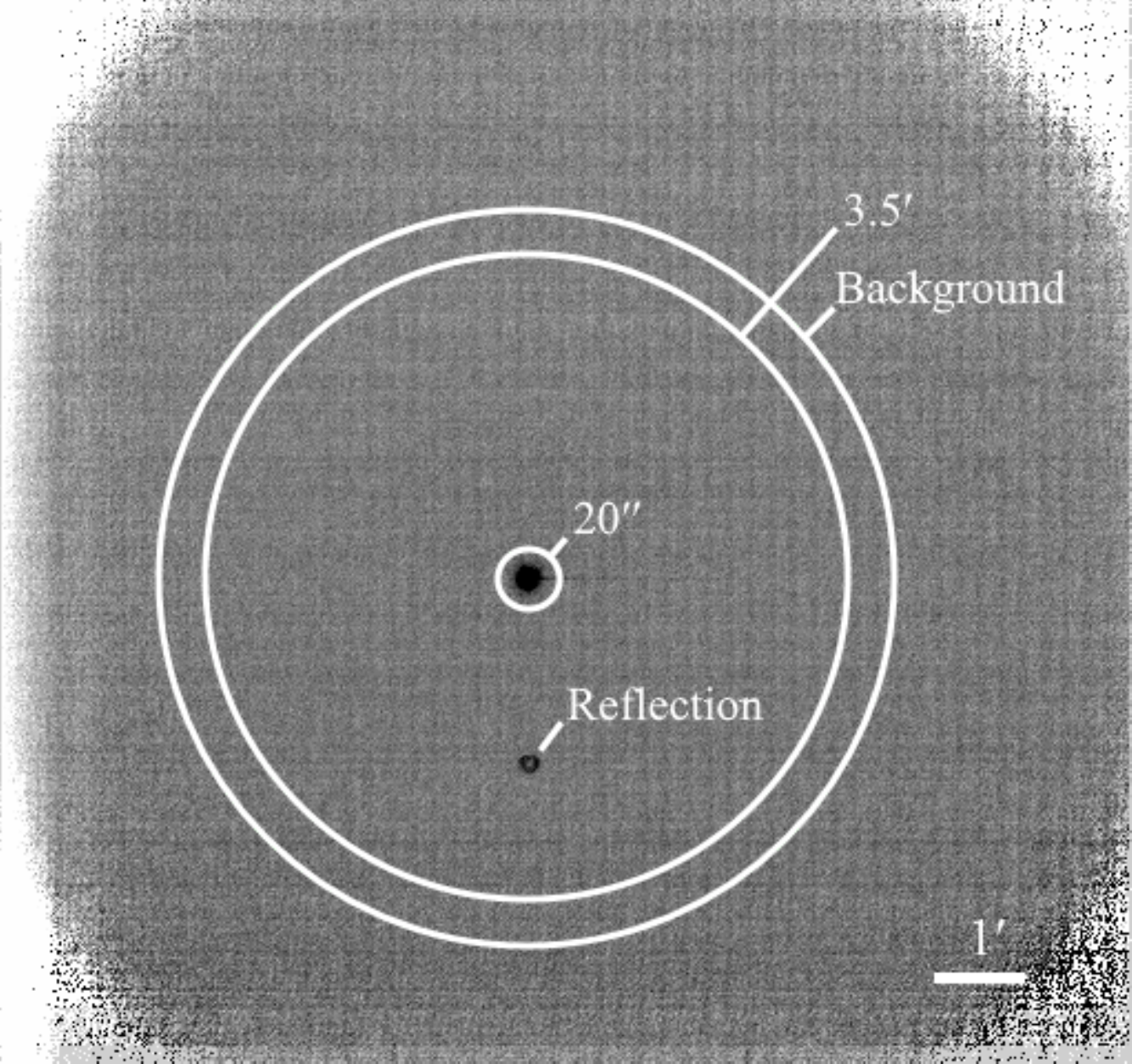}
\caption{Apertures used for the measurement of the stacked {\it BC} image. The image is displayed in a color-inverted linear scale. The marked inner circle, middle and outer annuli are used to measure nucleus, coma and background fluxes, respectively. The vignetting, visible at the borders and edges of the image, is caused by the dichroic and limits the maximum usable aperture to $\sim4.5'$ in radius. The marked ring-like artifact is caused by internal reflection due to the high brightness of Phaethon.}
\label{fig:bc_ap}
\end{figure}

\begin{figure}[!htb]
\centering
\includegraphics[width=0.9\textwidth]{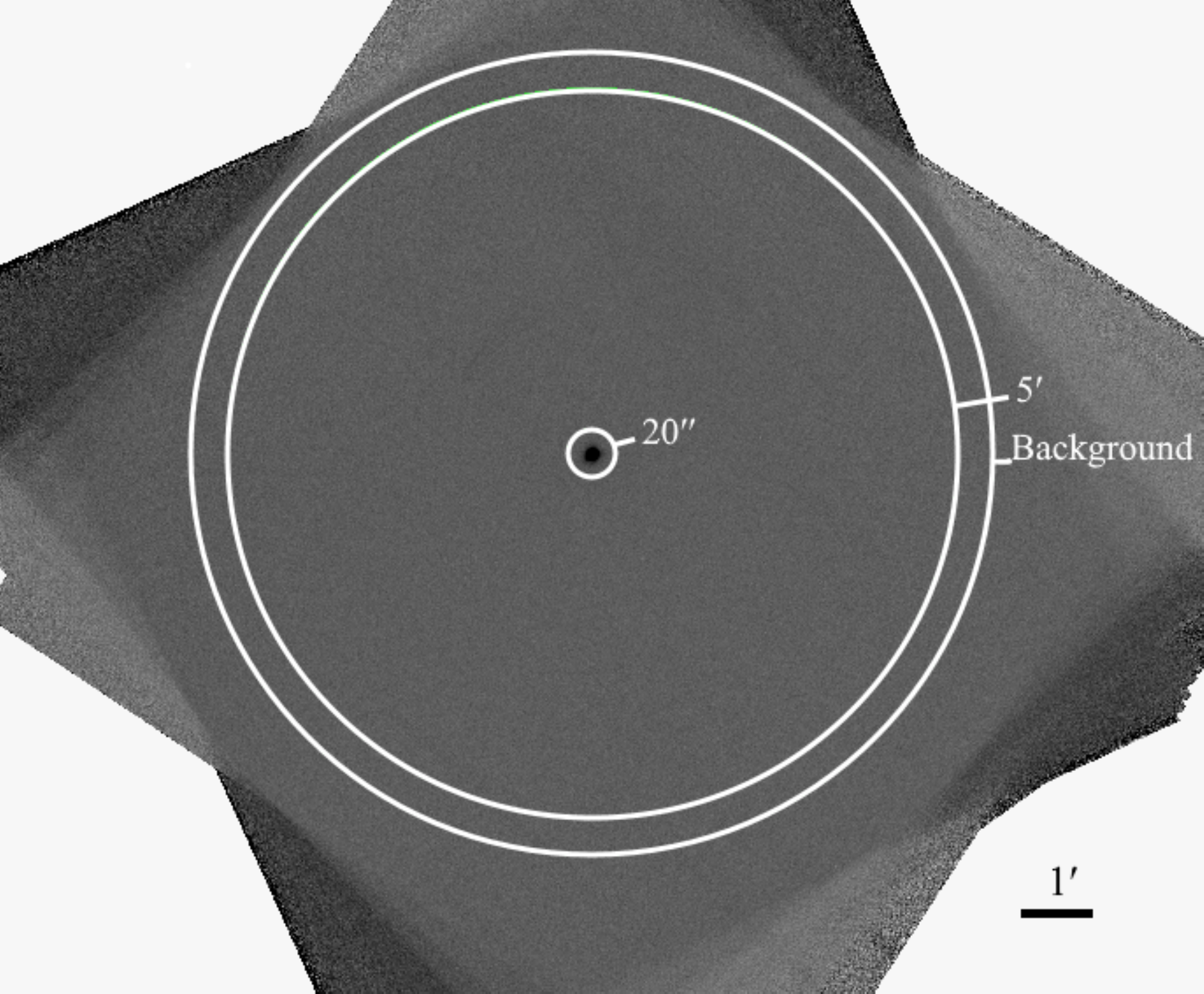}
\caption{Apertures used for the measurement of the stacked {\it r'} image. The image is displayed in color-inverted linear scale. The marked inner circle, middle and outer annuli are used to measure nucleus, coma and background fluxes, respectively. The vignetting caused by image dithering and changing sky conditions limits the maximum usable aperture to $\sim5'$ in radius.}
\label{fig:r_ap}
\end{figure}

\begin{figure}[!htb]
\centering
\includegraphics[width=0.5\textwidth]{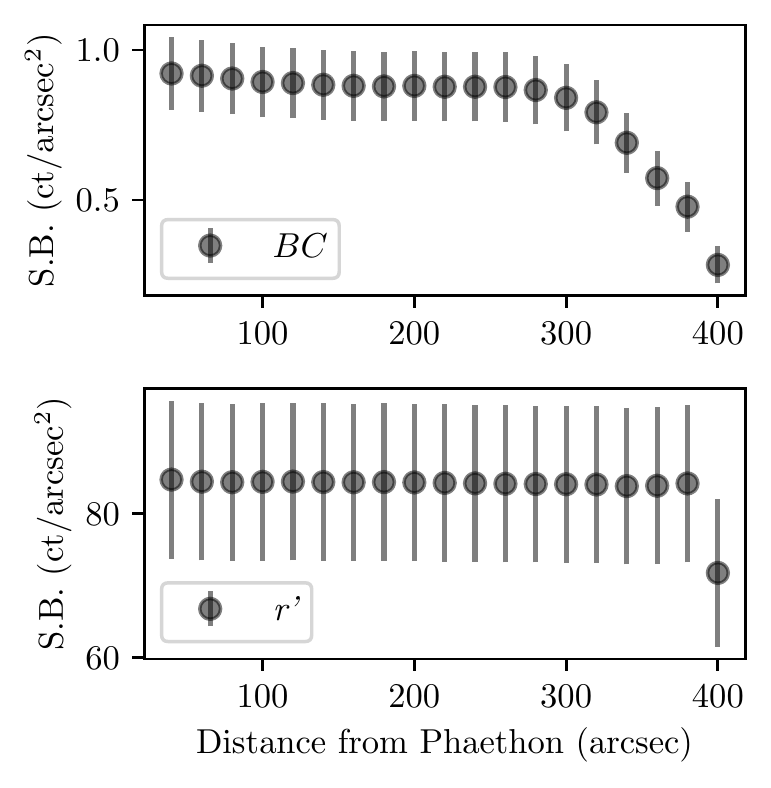}
\caption{Surface brightness (S.B., the total flux within each aperture divided by the area of the aperture) of a series of $20''$-wide annuli at increasing distances to Phaethon for {\it BC} and {\it r'} images. The fall-off within $100''$ in the {\it BC} curve is not statistically significant and is caused by the CCD readout tail (as seen in panels A-C in Figure~\ref{fig:img}). This feature is not present in the {\it r'} images as the sky was rotating during the observations to maintain the same parallactic angle for NIHTS. The steep sections beyond $\sim300''$ for {\it BC} and $\sim400''$ for {\it r'} are caused by the vignetting at image edges due to the use of the dichroic ({\it BC} only) as well as artifacts arising from image dithering.}
\label{fig:aper_rad}
\end{figure}

Apparently, an aperture radius of $5'$ ($\sim$20,000~km at Phaethon) would cover every possible scenario. However, as shown in Figures~\ref{fig:bc_ap}, \ref{fig:r_ap} and \ref{fig:aper_rad}, this isn't possible for the {\it BC} images owing to the vignetting caused by the dichroic as well as the dithering of the images, which only permits an aperture of up to $\sim4.5'$ in radius, potentially affecting the scenario where typical water-ice sublimation is the underlying ejection mechanism, though we note that this scenario is unlikely since Phaethon is clearly incompatible with typical comets. The {\it r'} filter does not have the vignetting issue, and can accommodate apertures as permitted by the camera, which is $\sim6'$ in radius. Hence, we use a $3.5'$ radius aperture for the {\it BC} images and a $5'$ radius aperture for the {\it r'} images, leaving a $0.5'$ annulus immediately outside the main aperture for background estimation, as shown in Figures~\ref{fig:bc_ap} and \ref{fig:r_ap}. We obtain $m_{BC}=11.16\pm0.02$ and $m_r=11.36\pm0.03$ where $m_{\lambda}$ is the magnitude for the specific band $\lambda$, with the uncertainties propagated from a combination of photometric uncertainties and background Gaussian noise. As we will show below, the latter assumption is problematic, but the statistical errors in catalogs are comparable in magnitude to the background noise, therefore this issue does not affect our results in a significant way.

The cross-section area of the dust can then be calculated by:

\begin{equation}
    C = \frac{\pi r_\mathrm{H}^2 \varDelta^2}{p_{\lambda} (1~\mathrm{au})^2 \phi(\alpha)} 10^{-0.4(m_{\lambda}-m_{\odot, \lambda})}
\end{equation}

\noindent with the uncertainty calculated by:

\begin{equation}
    [\Delta C] = 0.16 \ln{10} (m_{\lambda}-m_{\odot, \lambda}) [\Delta m_{\lambda}] C 
\end{equation}

\noindent where $p_{\lambda}=0.12$ is the geometric albedo in {\it BC} and {\it r'}, assumed to equal the general optical value \citep{Hanus2016}, $\phi(\alpha)$ is the phase function at phase angle $\alpha$ which can be calculated using the measurement made by \citet{Tabeshian2019} using $B$-band for {\it BC} and $R$-band for {\it r'}, $\Delta m_{\lambda}$ is the uncertainty of $m_{\lambda}$, and $m_{\odot, \lambda}$ is the apparent brightness of the Sun ($-26.29$ for {\it BC} and $-26.93$ for {\it r'}) derived from the color terms measured by \citet{Farnham2000} and \citet{Willmer2018}. We note that the choice between asteroidal and cometary phase function is important here. If we use a cometary phase function such as the one described by \citet{Marcus2007}, our final $C$ will be $2$--$10\times$ smaller, with increasing differences at larger $\alpha$ due to forward scattering enhancement of cometary dust. Although a dust coma from Phaethon will likely be subjected to the same forward scattering enhancement, our measurement here is dominated by the nucleus signal from Phaethon, therefore using the phase function measured for Phaethon is more appropriate. Additionally, as will be shown below, no flux excess that indicates the presence of a coma is detected. Hence, we conclude that the contribution of the phase function in the comet regime in the final derived numbers is negligible.

By inserting all these numbers, we derive $C_{BC}=29.7\pm3.4~\mathrm{km^2}$ and $C_{r}=41\pm14~\mathrm{km^2}$. The cross-section of Phaethon itself is somewhat uncertain, owing to the uncertainty in its shape and surface property \citep{Hanus2016, Taylor2019, Ye2019b, Devogele2020}. If we take the radius of Phaethon to be $\sim3$~km, the likely value suggested by radar measurement \citep{Taylor2019}, the cross-section of Phaethon is $C_\mathrm{Phaethon}=\pi R_\mathrm{Phaethon}^2=28.3~\mathrm{km^2}$, in agreement with the derived cross-section within the uncertainty, confirming the lack of dust excess as found in Figure~\ref{fig:profile}. On the other end, the cross-section of the nucleus could be as small as $20.4\pm1.6~\mathrm{km^2}$ based on the smaller nucleus derived from thermalphysical modeling \citep{Hanus2016}. This will increase our detection to $\sim2.4\sigma$, but still does not constitute a strong argument for the presence of a coma.

We note that the statistical uncertainty of the HB and PANSTARRS catalogs is about $1\%$--$1.5\%$ \citep{Farnham2000, Magnier2016}, comparable to the derived photometric uncertainties of Phaethon on an order-of-magnitude level. Is the uncertainty of the derived photometry dominated by catalog uncertainty? If so, can we further improve our results by doing differential photometry instead of absolute photometry?

To investigate this issue, we compare the raw counts derived from the earlier $3.5'$ (for {\it BC}) and $5'$ (for {\it r'}) apertures to that derived from a smaller, $20''$-radius aperture that only includes the nucleus in order to search for any excess in the $3.5'$/$5'$ apertures (as shown in Figures~\ref{fig:bc_ap} and \ref{fig:r_ap}). This eliminates the absolute calibration in the process and will improve the precision if the catalog uncertainty dominates the final uncertainty. The $20''$ radius is chosen based on the location where the radial brightness profiles in Figure~\ref{fig:profile} reach asymptotes. The counts are background-subtracted and are sigma-clipped at the $3\sigma$ level to reject artifacts, such as the internal reflection shown in Figure~\ref{fig:bc_ap}. We confirm through visual inspection that no possible coma signal is rejected by the sigma-clipping.

The relative brightness between the $3.5'$/$5'$ apertures and the ``nucleus'' apertures is found to be $(1.050\pm0.009)\mathcal{F}_\mathrm{nucleus}$ for {\it BC}, and $(1.128\pm0.088)\mathcal{F}_\mathrm{nucleus}$ for {\it r'}, where $\mathcal{F}_\mathrm{nucleus}$ is the apparent brightness of the nucleus. In an ideal case, an inactive nucleus should produce a relative photometry of unity; however, here our results show a nominal detection at $5.6\sigma$ (for {\it BC}) and $1.5\sigma$ (for {\it r'}) levels! To validate this result, we repeat the same procedure on the comparison stars defined in \S~\ref{sect:dust:profile}, and obtain $(1.083\pm0.018)\mathcal{F}_\mathrm{nucleus}$ for {\it BC} and $(1.079\pm0.034)\mathcal{F}_\mathrm{nucleus}$ for {\it r'} (4.6 and 2.3$\sigma$, respectively). Since the comparison stars do not have physical coma, the validation result suggests that the uncertainty is underestimated by a factor of several. Additional tests with simulated Gaussian-dominated noise images show that the stacked images have some remaining systematics as opposed to being Gaussian-noise-dominated as assumed in our error estimation. This can be caused by a variety of reasons, such as non-ideal flat-fielding, artifacts from optical scattering, weak bias pattern that isn't completely removed, and so on. Owing to the limited amount of data, we do not attempt to construct a background model that accounts for the correlation of background pixels, but instead simply estimate the true uncertainty based on the images of comparison stars. We estimate that the upper limit of dust excess is about $10\%$--$15\%$ of the nucleus flux, grossly comparable to the {\it BC} value derived from absolute photometry and represents a $3\times$ improvement to the {\it r'} value. This shows that the true background noise contributes a similar amount (or slightly less, for the case of {\it r'}) of error to the catalog error. Therefore, we conclude that differential photometry does not provide much improvement over absolute photometry when the statistical uncertainty of the catalog is $\lesssim2\%$.

\subsubsection{Dust Production Rate}

The upper limit of the dust production rate can be approximated by

\begin{equation}
    \dot{M} \sim \frac{\rho_\mathrm{d} [\Delta C] \bar{a} v_\mathrm{ej}}{\theta \varDelta}
\end{equation}

\noindent where $\rho_\mathrm{d}=2600~\mathrm{kg~m^{-3}}$ is the bulk density of the grain \citep{Borovicka2019}, $\theta$ is the aperture radius in radians, $\bar{a}=1~\micron$ is the characteristic grain size for optical observations, and all other variables follow the definitions given above. The value of $v_\mathrm{ej}$ is most uncertain, and can vary between $3$--$100~\mathrm{m~s^{-1}}$ depending on the ejection mechanism, with $3~\mathrm{m~s^{-1}}$ being the gravitational escape ejection and $100~\mathrm{m~s^{-1}}$ being the upper bound of the sublimation-driven ejection. By inserting the numbers derived in \S~\ref{sect:dust:aper} with $\Delta C\sim4~\mathrm{km^2}$ and a $5'$ aperture radius following the discussions in the same section, we obtain a $3\sigma$ upper limit of dust product rate of $0.005$--$0.2~\mathrm{kg~s^{-1}}$. This is grossly in line with \citet{Hsieh2005} and \citet{Tabeshian2019}, who derived $\dot{M}\lesssim0.01~\mathrm{kg~s^{-1}}$ and $\dot{M}\lesssim0.06$--$0.2~\mathrm{kg~s^{-1}}$, respectively, though we note that they have each adopted a different set of parameters: \citet{Hsieh2005} used $\rho_\mathrm{d}=1000~\mathrm{kg~m^{-3}}$, $\bar{a}=0.5~\micron$, and an equivalent $v_\mathrm{ej}=30~\mathrm{m~s^{-1}}$, and \citet{Tabeshian2019} used $\rho_\mathrm{d}=3000~\mathrm{kg~m^{-3}}$, $\bar{a}=1~\micron$, and an equivalent $v_\mathrm{ej}\sim500~\mathrm{m~s^{-1}}$. If we consider Phaethon as a nearly extinct active body which likely has a lower $v_\mathrm{ej}$, and adopt $v_\mathrm{ej}\sim10~\mathrm{m~s^{-1}}$, we would obtain $\dot{M}\lesssim10^{-2}~\mathrm{kg~s^{-1}}$ using the measurements reported in all three studies.

\subsection{Gas}

We use two approaches to constrain Phaethon's gas production rate: imaging with a narrowband gas filter and spectroscopy. The narrow-band imaging permits measurement over a large sky area, which is particularly advantageous due to Phaethon's close distance to the Earth; while spectroscopy permits simultaneous measurements of multiple gas species and provides an unambiguous identification if a signal is detectable.

\subsubsection{Narrowband CN Imaging}
\label{sect:cn}

\begin{figure*}[!htb]
\centering
\includegraphics[width=0.7\textwidth]{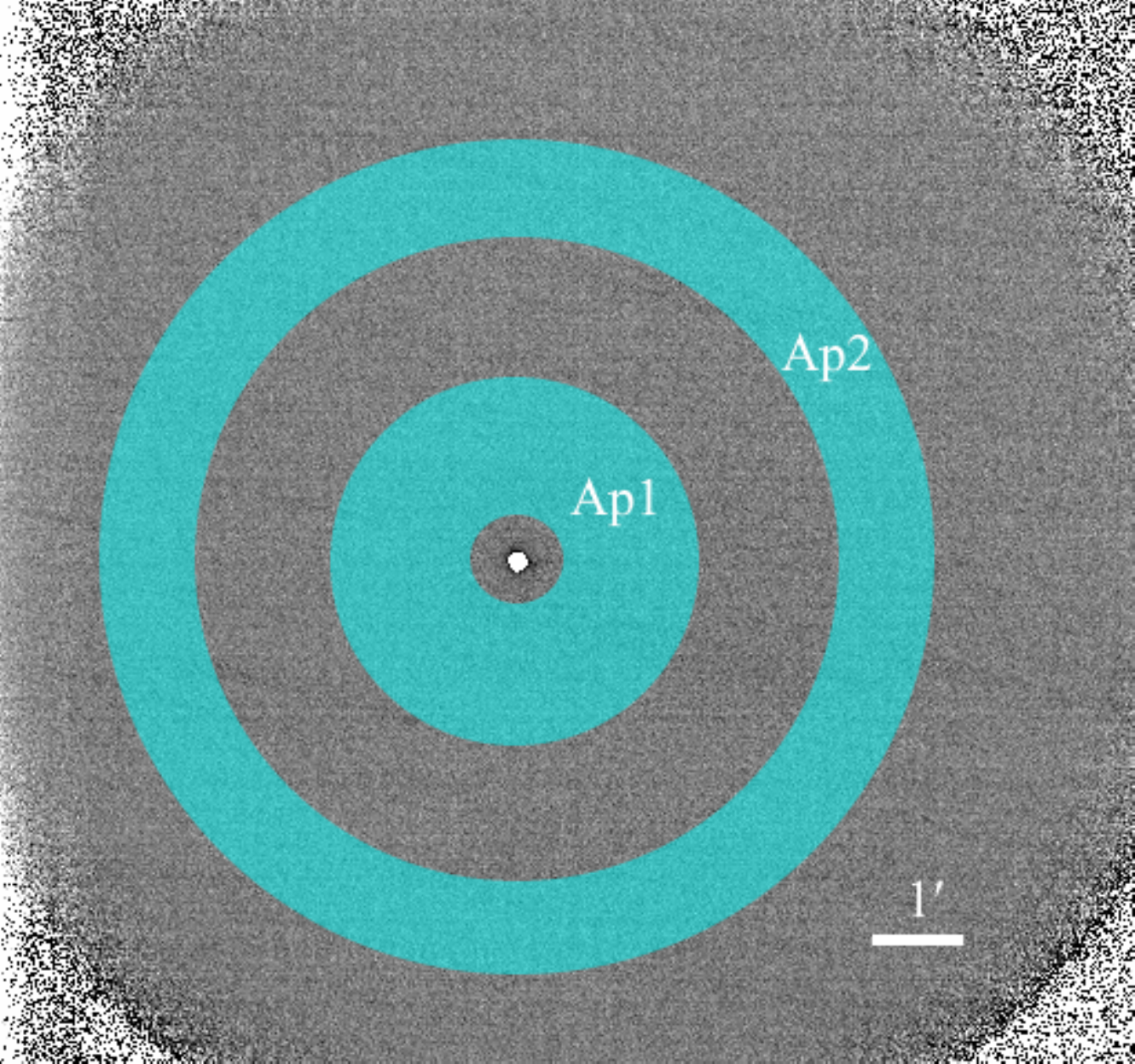}
\caption{Two annular apertures (denoted as {\tt Ap1} and {\tt Ap2}) used for the measurement of CN flux. The underlying image is the color-inverted, median-stacked CN image with sigma-clipping rejection at $3\sigma$ level. The masked pixels are displayed in white.}
\label{fig:cn-ap}
\end{figure*}

To minimize the contamination from background star trails, we first reprocess the {\it CN} images with a sigma-clipping rejection of background sources at $3\sigma$ level, and then median combine them into one stacked image, shown as the underlying image in Figure~\ref{fig:cn-ap}.

Given Phaethon's close distance to the Earth during the approach, a hypothetical gas coma would be a few degrees in size and is therefore much wider than LMI's field-of-view, presenting a challenge for the determination of the \textit{true} sky background and any residual instrumental bias. Hence, we do not attempt to measure the sky background, but instead measure the integrated flux of two different-sized annular apertures centered on Phaethon, and then use the difference between the two to derive the CN emission. This is simply because the column density of the gas species decreases as the projected distance to the nucleus increases \citep[e.g.][Figure~1]{Combi2004}, with the amount calculable with a \citet{Haser1957} model. The configuration of the two apertures is depicted in Figure~\ref{fig:cn-ap}: the inner annulus, {\tt Ap1}, is between $0.5'$--$2'$ from Phaethon, while the outer annulus, {\tt Ap2}, is between $3.5'$--$4.5'$ from Phaethon. 

\begin{figure}[!htb]
\centering
\includegraphics[width=0.5\textwidth]{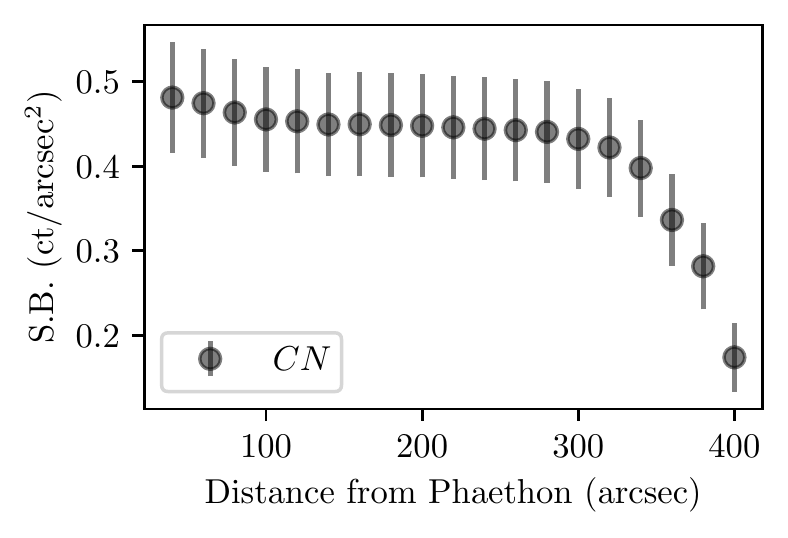}
\caption{Surface brightness (S.B.) of a series of $20''$-wide annuli at increasing distances to Phaethon for {\it CN} images. The bump within $100''$ is not statistically significant and is caused by the CCD readout tail. The steep section beyond $\sim300''$ is caused by vignetting at the image edges.}
\label{fig:aper_rad_cn}
\end{figure}

Based on the findings in \S~\ref{sect:dust}, we assume the underlying dust continuum in the coma region is negligible. The inner radius of {\tt Ap1} is selected to ensure that the continuum from Phaethon itself is excluded. While it is possible to subtract the continuum flux inside the inner aperture using the {\it BC} images and the spectral slope derivable from the spectrum described in \S~\ref{sect:obs:spec}, this introduces more uncertainty into the process since any CN signal inside the inner aperture is very weak and the continuum from Phaethon's nucleus will dominate. Furthermore, the Haser model predicts that the {\it CN } flux within a $30''$ radius aperture only contributes $\sim10\%$ of the total CN flux within an aperture of radius $2'$ (the outer radius of {\tt Ap1}) due to the much larger collecting area of the latter, so the inclusion of the innermost CN component does not significantly improve the signal in {\tt Ap1}.  The outer radius of {\tt Ap2} is determined by a background curve-of-growth analysis similar to one conducted for {\it BC} and {\it r'} images (i.e. Figure~\ref{fig:aper_rad}), which shows a usable field of $\sim4.5'$ in radius (Figure~\ref{fig:aper_rad_cn}). The annuli are placed as far apart from each other as possible to maximize their flux difference.

The gas production rate $Q_i$ of species $i$ can then be calculated from the flux of this species $\mathcal{F}_i$ by

\begin{equation}
\label{eq:q}
    Q_i = \frac{4 \pi \varDelta^2 \mathcal{F}_i}{g(i, r_\mathrm{H}) \tau(i, r_\mathrm{H}) \mathcal{H}_i}
\end{equation}

\noindent where $g(i, r_\mathrm{H})=g(\mathrm{i, 1~au}) r_\mathrm{H}^{-2}$ is the fluorescence efficiency (or the ``g-factor''), $\tau(i, r_\mathrm{H})=\tau(\mathrm{i, 1~au}) r_\mathrm{H}^{2}$ is the lifetime of this species, and $\mathcal{H}_i$ is the collection efficiency for this species (the fraction of gas molecules in the observing aperture as compared to an infinitely large aperture; also called Haser correction) to be solved by the Haser model. Although we do not know the absolute values of each $\mathcal{F}_i$ in \texttt{Ap1} and \texttt{Ap2}, we know how they relate to each other, which allows us to iteratively solve $Q_i$.

The Haser model requires the knowledge of the scale lengths of parent and daughter species, as well as the radial outflow speed of the gas, $v_i$. The scale lengths of several common cometary species have been estimated based on the observations of bright comets \citep[e.g.][]{Randall1992, Ahearn1995}, which we tabulate in Table~\ref{tbl:spec}. The outflow speed is typically taken from the vectorial model which assumes a constant speed of $\sim1$~km/s at all heliocentric distances \citep[e.g.][]{Festou1981, Festou1981b, Festou1985, Feldman1987}. However, later studies have disputed this assumption \citep[e.g.][]{BockeleeMorvan1990, TacconiGarman1990}, suggesting that the outflow speed is, in fact, dependent on heliocentric distance and gas production rates, though the exact relationship is not clear. \citet{Delsemme1982} compiled the works of others and noted a $\propto Q_i^{1/2}$ dependence, which were later confirmed by others \citep[e.g.][]{Cochran1993}, presumably due to the collisional process between molecules in these comets, though the data only included comets with $Q_i > 10^{28}~\mathrm{molecule~s^{-1}}$ and the relation may not extend to weakly active comets. \citet{Cochran1993} and \citet{Budzien1994} both concluded that a satisfactory general value is $v_i = 0.85r_\mathrm{H}^{-1/2}$~km~s$^{-1}$, and this relation has been frequently adopted \citep[e.g., ][]{Combi2004,Feldman2018}. On the other hand, \citet{BockeleeMorvan1990} and \citet{Tseng2007} analyzed observations of distant, weakly active comets up to 4.5~au from the Sun, and suggested $v_i=0.5$--$0.6~\mathrm{km~s^{-1}}$ for distant weakly active comets, again consistent with a simple $\propto r_\mathrm{H}^{-1/2}$ [km/s] law. Hence, for simplicity, we consider a single, nominal model with $v_i(r_\mathrm{H})=(r_\mathrm{H}/\mathrm{1~au})^{-1/2}~\mathrm{[km/s]}$, but caution that this may not be representative of extinct or near-extinct objects such as Phaethon.

We then proceed to derive the relation between the aperture-corrected flux ``excess'' of {\tt Ap1} relative to {\tt Ap2}, described as

\begin{equation}
\label{eq:delta_fi}
    \Delta \mathcal{F}_i = \mathcal{F}_{i, \tt Ap1} - \left( \frac{\mathcal{A}_{\tt Ap1}}{\mathcal{A}_{\tt Ap2}} \right) \mathcal{F}_{i, \tt Ap2}
\end{equation}

\noindent where $\mathcal{F}_{i, \tt Ap1}$, $\mathcal{F}_{i, \tt Ap2}$ is the flux in {\tt Ap1} and {\tt Ap2}, and $\mathcal{A}_{\tt Ap1}$, $\mathcal{A}_{\tt Ap2}$ is the on-sky area of {\tt Ap1} and {\tt Ap2}, respectively (with annuli dimensions specified above). We then measure $\Delta \mathcal{F}_i$ from the stacked image, using the parameters published in \citet{Ahearn1995} and \citet{Schleicher2010} (tabulated in Table~\ref{tbl:spec}) as appropriate for CN to convert the solved column densities to observed flux. The fluorescence efficiencies calculated by \citet{Schleicher2010} accounted for the impact from the Swings effect, the variation of the relative intensities of the CN lines due to the Doppler shift of the solar Franhofer lines with respect to the comet's CN emissions \citep{Swings1941}. We obtain $\Delta \mathcal{F}_i = (8.5\pm0.4)\times10^{-15}~\mathrm{erg~\AA^{-1}~cm^{-2}~s^{-2}}$, showing a detection at $21\sigma$! As a validation check, we repeat the same process for the images of the flux standard HD 37112 (used as a comparison star due to its similar airmass to Phaethon at the time of the observation), and obtain $\Delta \mathcal{F}_i = (6.6\pm0.2)\times10^{-15}~\mathrm{erg~\AA^{-1}~cm^{-2}~s^{-2}}$, similarly showing a high sigma level. Similar to the situation we encountered in \S~\ref{sect:dust:aper}, this is likely caused by correlated background noises, and do not indicate the presence of gas coma in the {\it CN} images. Again, due to the small amount of data, we do not attempt to model the background, but instead estimate the upper limit of $\Delta \mathcal{F}_i$ to be $\sim1\times10^{-14}~\mathrm{erg~\AA^{-1}~cm^{-2}~s^{-2}}$.

To express $Q_i$ as a function of measurable $\Delta \mathcal{F}_i$, we first rearrange Equation~\ref{eq:q} as:

\begin{equation*}
    F_i = \frac{Q_i g(i, r_\mathrm{H}) \tau(i, r_\mathrm{H}) \mathcal{H}_i}{4 \pi \varDelta^2}
\end{equation*}

Replacing $\mathcal{F}_{i, \tt Ap1}$ and $\mathcal{F}_{i, \tt Ap2}$ in Equation~\ref{eq:delta_fi}, we will have:

\begin{equation*}
    \Delta F_i = \frac{Q_i g(i, r_\mathrm{H}) \tau(i, r_\mathrm{H})}{4 \pi \varDelta^2} \left[ \mathcal{H}_{i, \tt Ap1} - \left( \frac{\mathcal{A}_{\tt Ap1}}{\mathcal{A}_{\tt Ap2}} \right) \mathcal{H}_{i, \tt Ap2} \right]
\end{equation*}

\noindent which can be rearranged into

\begin{equation}
    Q_i = \frac{4 \pi \varDelta^2 [\Delta \mathcal{F}_i]}{g(i, r_\mathrm{H}) \tau(i, r_\mathrm{H})} \left[ \mathcal{H}_{i, \tt Ap1} - \left( \frac{\mathcal{A}_{\tt Ap1}}{\mathcal{A}_{\tt Ap2}} \right) \mathcal{H}_{i, \tt Ap2} \right]^{-1}
\end{equation}

\noindent where $\mathcal{H}_{i, \tt Ap1}$, $\mathcal{H}_{i, \tt Ap1}$ are the collection efficiency (i.e. Haser correction) for {\tt Ap1} and {\tt Ap2}, respectively. This equation is solved by the Haser model provided by \texttt{sbpy} \citep{Mommert2019}. For reference, the predicted ratio of the total CN signal between {\tt Ap1} and {\tt Ap2} is $\sim0.93:1$ regardless of $Q_\mathrm{CN}$. By using the upper limit of $\Delta \mathcal{F}_i$ we just derived, we solve the upper limit of the CN gas to be $1.4\times10^{22}~\mathrm{molecule~s^{-1}}$.

\subsubsection{Spectroscopy}
\label{sect:spec}

\begin{figure*}[!htb]
\centering
\includegraphics[width=\textwidth]{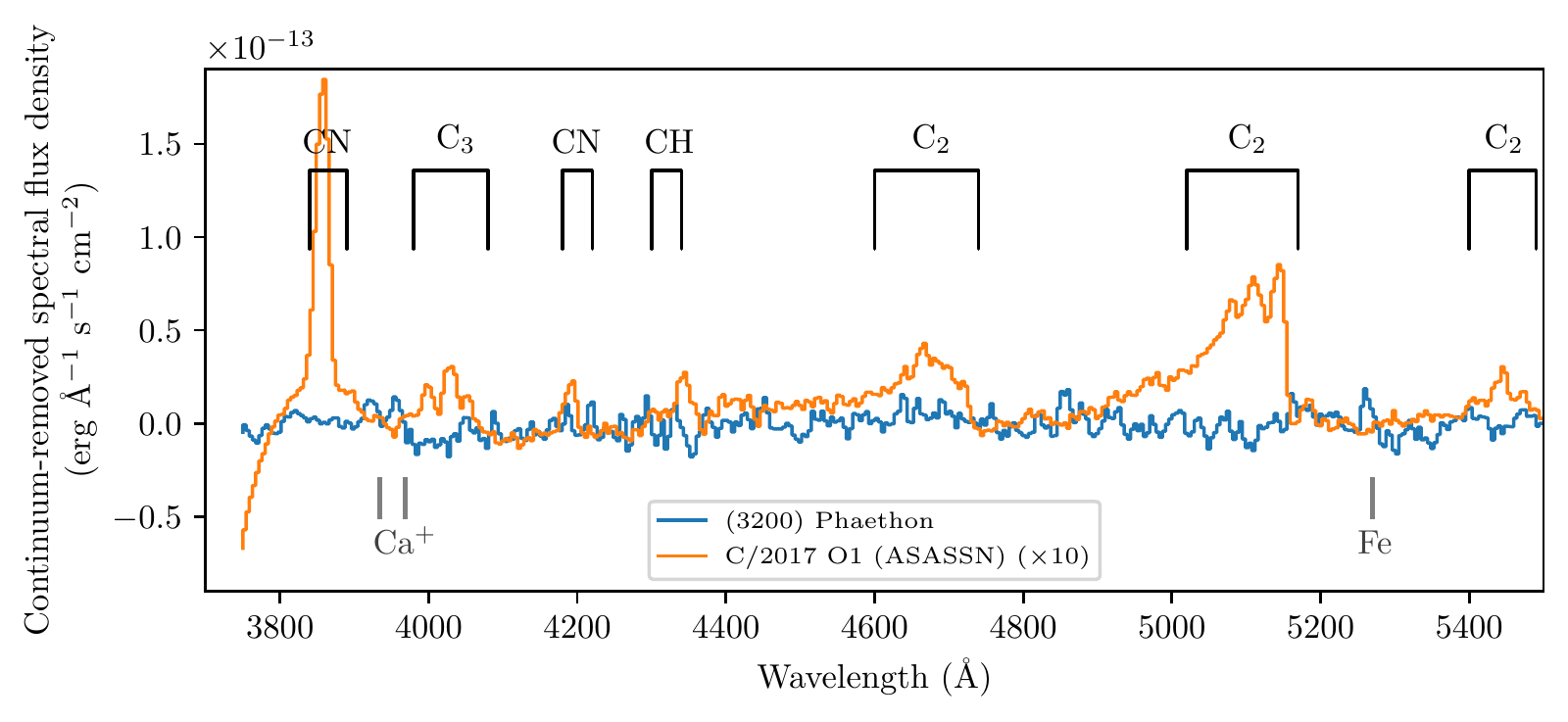}
\caption{Normalized, continuum-removed spectrum of Phaethon as observed with LDT/DeVeny spectrograph on 2017 December 15, with major cometary emission lines and the Fraunhofer artifacts (the oversubtracted Ca$^+$ and Fe lines from the solar analog) marked. The spectrum of comet C/2017 O1 (ASASSN), observed with LDT/DeVeny on the same night and multiplied by a factor of 10 for clarity, is also shown as a reference of the major cometary emission lines. The spectra are aligned to the continuum points. Cometary emission lines are labeled using the lines listed in \citet{Brown1996} and \citet{Farnham2000}.}
\label{fig:spec}
\end{figure*}

To identify possible gas emissions and measure their rates, we remove the continuum in Phaethon's spectrum using a color-corrected solar spectrum, where the spectral color is derived by fitting a 3rd-order Chebyshev polynomial to Phaethon's spectrum at the gas-free continuum points suggested by \citet{Farnham2000}. We then repeat the same procedures for C/ASASSN's spectrum. These procedures are done with the {\tt specutils} package \citep{Specutils2019}. The resulting normalized spectra of Phaethon and C/ASASSN, shown in Figure~\ref{fig:spec}, show no cometary emission from Phaethon. Similar to the analysis of the narrowband images, the derivation of the absolute flux and gas production rate from the spectrum requires a non-standard treatment, since the sky aperture can be contaminated by gas. We follow the steps laid out in \S~\ref{sect:cn} and use the flux excess between the on-target aperture and (contaminated) ``sky'' to iteratively solve the gas production rate, with the various apertures defined in \S~\ref{sect:obs:spec} accounted for. The constraints are derived from the $3\sigma$ upper limits of the fluxes integrated over the bandpasses of each species \citep[c.f.][Figure~4]{Cochran2012}. Again, following the earlier findings, we assume the underlying dust continuum is negligible. The results, as well as the spectroscopic constants used, are tabulated in Table~\ref{tbl:spec}.

\begin{deluxetable*}{ccccccc}
\tablecaption{Summary of spectroscopic constants, measured flux excess, and gas rates for each cometary species. The spectroscopic constants are quoted from \citet{Schleicher2010} for CN with the Swings effect corrected, and \citet[][Table II]{Ahearn1995} for other species. \label{tbl:spec}}
\tabletypesize{\scriptsize}
\tablehead{
\colhead{Species $i$} & \colhead{Window\tablenotemark{\scriptsize{a}}} & \colhead{$g(i, \mathrm{1~au})$} & \multicolumn{2}{c}{Haser scale length at 1~au} & \colhead{$\Delta \mathcal{F}_i$ ($3\sigma$ limit)} & \colhead{$Q_i$ ($3\sigma$ limit)} \\
 \cmidrule{4-5}
  \colhead{} & \colhead{} & \colhead{} & \colhead{Parent} & \colhead{Daughter} & \colhead{} & \\
\colhead{} & \colhead{(\AA)} & \colhead{($\mathrm{erg~s^{-1}}$)} & \colhead{(km)} & \colhead{(km)} & \colhead{($\mathrm{erg~cm^{-2}~s^{-1}~\AA^{-1}}$)} & \colhead{($\mathrm{molecule~s^{-1}}$)}
} 
\startdata
CN (spec.) & $3869\pm30$ & $3.2\times10^{-13}$\tablenotemark{\scriptsize{b}} & $1.3\times10^4$ & $2.1\times10^5$ & $<6.0\times10^{-15}$ & $<6.1\times10^{24}$ \\
CN (im.) & $3870\pm31$ & '' & '' & '' & See \S~\ref{sect:cn} & $<2.3\times10^{22}$ \\
C$_3$ (spec.) & $4062\pm30$ & $1\times10^{-12}$ & $2.8\times10^3$ & $2.7\times10^4$ & $<1.2\times10^{-14}$ & $<7.9\times10^{23}$ \\
C$_2$ (spec.) & 4850--5150 & $4.5\times10^{-13}$ & $2.2\times10^4$ & $6.6\times10^4$ & $<1.9\times10^{-14}$ & $<4.0\times10^{25}$ \\
\enddata
\tablenotetext{a}{This column has different meaning for different techniques: for narrowband \textit{CN} imaging, this refers to the bandpass of the filter; for spectroscopy, this refers to the aperture used for signal measurement.}
\tablenotetext{b}{This value corresponds to the $g$-factor of an object with $r_\mathrm{H}=1$~au and heliocentric speed $\dot{r_\mathrm{H}}=-28$~km/s as appropriate for Phaethon at the time of the observation. See \citet{Schleicher2010}.}
\end{deluxetable*}

\subsection{Fragment Search}
\label{sect:frag}

We visually inspected the stacked broadband images, shown in Figure~\ref{fig:img}c and d, for fragments in Phaethon's vicinity, but did not find any. The $5\sigma$ limiting magnitude of the combined $8\times300$~s image (Figure~\ref{fig:img}c) is $r'=24.0$, equivalent to a 15~m body at Phaethon's distance, assuming a Phaethon-like albedo of $p_V=0.12$ \citep{Hanus2016}. However, we caution that these images are heavily affected by star trails and a saturated Phaethon; additionally, any hypothetical fragment will be situated along the $\pm v$ vector (unless it has an unrealistically high out-of-plane ejection speed), which is only at a small angle with the background star trails, and almost coincides with Phaethon's readout trail. Therefore this very deep search is by no means complete. On the other hand, the $5\sigma$ limit of the combined unsaturated image shown in Figure~\ref{fig:img}d is $r'=20.4$~mag, equivalent to a 100~m body at Phaethon's distance assuming a Phaethon-like albedo. These images are not as deep as the 300~s images, but are free of background contamination. Our negative result is in line with previous reports as well as observations conducted during the same close approach, including ground- and space-based optical searches \citep{Jewitt2018, Ye2018, Tabeshian2019}, thermal \citep{Jewitt2019}, and radar observations \citep{Taylor2019}.

\section{Discussion}
\label{sec:disc}

\subsection{Comet Or Asteroid?}

Since the discovery of Phaethon in 1983, there has been a debate on its true nature, though now most, if not all the measured properties seem to indicate it being more asteroid-like than comet-like. Apart from the connection to the Geminid stream, of which the formation mechanism remains unknown, Phaethon's only evidence of mass-loss -- its near-Sun activity -- is clearly incompatible with typical comets \citep{Jewitt2010, Hui2017}. The derived upper limit for CN, presented in \S~\ref{sect:cn}, is nearly an order of magnitude improvement over the previous best result by \citet{Chamberlin1996}, and is on par with the lowest numbers ever measured/constrained for comets \citep[133P/Elst--Pizarro had $Q(CN)<1.3\times10^{21}~\mathrm{molecule~s^{-1}}$, and P/2016 BA14 (PANSTARRS) had $Q(CN)=1.4\times10^{22}~\mathrm{molecule~s^{-1}}$, c.f.][]{Licandro2011, Li2017}. The derived upper limit for dust emission is on par with the best limits derived previously \citep{Hsieh2005, Tabeshian2019}. How much further can we go?

Assuming a certain CN/OH ratio, we can use the CN rate to set an upper limit to the water-ice-driven active area on the surface of Phaethon. \citet{Ahearn1995} observed 41 comets and found the range of $\log{\mathrm{CN/OH}}$ to be $-2.94$ to $-2.17$. Applying this to Phaethon and using the ice sublimation model described by \citet{Cowan1979}\footnote{\url{https://pdssbn.astro.umd.edu/tools/ma-evap/index.shtml}.}, we derive an upper limit to the active area between 100--$4000~\mathrm{m^2}$, subjected to the exact $\log{\mathrm{CN/OH}}$ ratio or sublimation mode (subsolar or isothermal). This can be represented by an active area up to 35~m in radius and corresponds to a minuscule active fraction of $<0.003\%$, an order of magnitude smaller than the case of 209P/LINEAR, the lowest active fraction ever measured for a comet \citep[$0.03\%$;][]{Schleicher2016}. This result is consistent with the apparent lack of hydration features derived from near-infrared spectroscopy \citep{Takir2020}, as well as the predicted depletion of ices due to Phaethon's extreme orbit \citep{Jewitt2010, Yu2019}.

Could Phaethon's hypothetical comet-like activity be driven by a volatile species other than water ice? (3552) Don Quixote, for example, has a CO$_2$ coma that is comparable to other short-period comets in intensity \citep{Mommert2014, Mommert2020}. Certain comets are primarily driven by CO sublimation rather than the classic water ice sublimation, such as C/2016 R2 (PANSTARRS) \citep{Biver2018, Cochran2018, Wierzchos2018} and possibly 2I/Borisov \citep{Bodewits2020, Cordiner2020}. However, these lines are best to be observed from space due to the strong telluric absorption. So far, Phaethon's CO/CO$_2$ emissions remain largely unconstrained. The only reported constraint measured from submillimeter observation is $\sim6$ orders of magnitude higher than the CN upper limit \citep{Wiegert2008}. \citet{Mommert2020} found that the source that drives Don Quixote's activity can be represented as a single vent 45~m in radius, producing dust at an order of $\sim10^{-1}~\mathrm{kg~s^{-1}}$ assuming $\rho_\mathrm{d}=500~\mathrm{kg~m^{-3}}$, $\bar{a}=100~\micron$, and $v_\mathrm{ej}=10~\mathrm{m~s^{-1}}$. The active area and dust production of Phaethon is therefore even lower than an extreme case such as Don Quixote.

In conclusion, the constraints derived by this and other earlier works shows that Phaethon is an exceptional object even when being compared to the most extreme cases ever observed in comets.

\subsection{Implication for the {\it DESTINY$^+$} Mission}

The {\it DESTINY$^+$} mission to Phaethon, currently planned to launch in 2022, will carry a dust module and a spectrophotometric camera to study dust and gas ejection from the asteroid \citep{Arai2019, Kruger2019, Ishibashi2020}. The flyby is currently planned for 2026 August 1 at $r_\mathrm{H}=0.87$~au during Phaethon's inbound phase, in an orbital configuration similar to our observation (inbound at $r_\mathrm{H}=1.06$--0.99~au). Assuming Phaethon's behavior does not change appreciably from orbit to orbit, {\it DESTINY$^+$} will experience similar dust and gas rates as during our observations.

Our observations confirmed an upper limit of the dust production rate of $\sim10^{-2}~\mathrm{kg~s^{-1}}$ for micron-sized dust, or $\sim10^{-1}~\mathrm{kg~s^{-1}}$ for $100~\micron$-sized dust that is of more concern from a hazard perspective. Assuming an isotropic, sublimation-driven ejection from Phaethon and a $\sim20$ minutes flyby at a relative speed of 30~km/s and a distance of 500~km \citep{Arai2019, Kruger2019}, the 1-m-sized {\it DESTINY$^+$} spacecraft will encounter up to $10^5$ micron-sized particles \citep[or $10^3$ for the Dust Analyzer, confirming the number derived by][]{Masanori2018}, or $10^2$ submillimeter-sized dust. However, in the case of a lower ejection speed, the number of particle encountered can be higher: assuming a gravitational escape speed, the number of micron-sized particles can be up to $10^7$. We note that these numbers are only upper limits, since the dust rate derived from our observation is also a upper limit.

Although {\it DESTINY$^+$} will carry a multi-band spectrophotometric camera, our analysis suggests that direct detection of the gas coma during the flyby is unlikely. This is echoed by the non-detection of gas coma at comet 67P/Churyumov--Gerasimenko in the in-situ observation made by the {\it Rosetta} spacecraft \citep{Bodewits2016}. The best chance to detect a gas coma would be when Phaethon is close to the Sun with large observer--Phaethon distance; however, such observation would be difficult to be conducted from Earth's orbit due to Phaethon's extreme orbit.

\section{Conclusion} \label{sec:conc}

We conducted a deep search for gas and dust emissions from Phaethon using the 4.3~m Lowell Discovery Telescope, during Phaethon's remarkably close approach to the Earth in 2017 December. No signs of activity were detected. The $3\sigma$ upper limit of dust production rate is 0.007--0.2~$\mathrm{kg~s^{-1}}$; while the $3\sigma$ upper limit of CN gas production is $2.3\times10^{22}~\mathrm{molecule~s^{-1}}$. We did not find any 100-m-class fragment in Phaethon's vicinity; a deeper, but star-contaminated search also did not reveal any fragment down to 15~m range. Our results are consistent with the idea that Phaethon is now largely inactive, and that the Geminid meteoroid stream, presumably originated from Phaethon, must have been generated during the times when Phaethon was much more active.


\acknowledgments

We thank two anonymous referees for their careful reviews and useful comments, Heidi Larson and Teznie Pugh for operating the telescope, as well as Maxime Devog\`{e}le for discussion on the reduction of DeVeny data. QY, MMK and MSK are supported by NASA Near Earth Object Observations grant NNX17AK15G. NAM and AG recognize support from NASA NEOO grant NNX17AH06G. These results made use of the Lowell Discovery Telescope at Lowell Observatory. Lowell is a private, non-profit institution dedicated to astrophysical research and public appreciation of astronomy and operates the LDT in partnership with Boston University, the University of Maryland, the University of Toledo, Northern Arizona University and Yale University. The Large Monolithic Imager was built by Lowell Observatory using funds provided by the National Science Foundation (AST-1005313). The upgrade of the DeVeny optical spectrograph has been funded by a generous grant from John and Ginger Giovale and by a grant from the Mt. Cuba Astronomical Foundation.

The Pan-STARRS1 Surveys (PS1) and the PS1 public science archive have been made possible through contributions by the Institute for Astronomy, the University of Hawaii, the Pan-STARRS Project Office, the Max-Planck Society and its participating institutes, the Max Planck Institute for Astronomy, Heidelberg and the Max Planck Institute for Extraterrestrial Physics, Garching, The Johns Hopkins University, Durham University, the University of Edinburgh, the Queen's University Belfast, the Harvard-Smithsonian Center for Astrophysics, the Las Cumbres Observatory Global Telescope Network Incorporated, the National Central University of Taiwan, the Space Telescope Science Institute, the National Aeronautics and Space Administration under Grant No. NNX08AR22G issued through the Planetary Science Division of the NASA Science Mission Directorate, the National Science Foundation Grant No. AST-1238877, the University of Maryland, Eotvos Lorand University (ELTE), the Los Alamos National Laboratory, and the Gordon and Betty Moore Foundation.

\software{{\tt AstroImageJ} \citep{Collins2017}, {\tt ccdproc} \citep{Craig2015}, {\tt IRAF} \citep{Tody1986, Tody1993}, {\tt PHOTOMETRYPIPELINE} \citep{Mommert2017}, {\tt photutils} \citep{Bradley2016}, {\tt sbpy} \citep{Mommert2019}, {\tt specutils} \citep{Specutils2019}}
\facilities{LDT}

\end{CJK*}
\bibliographystyle{aasjournal}
\bibliography{ms}{}



\end{document}